# AGRICULTURAL INNOVATION IN THE INCA EMPIRE: A SUSTAINABLE APPROACH TO FOOD PRODUCTION

**Prof. Luis-Felipe Arizmendi**
Senior Lecturer
Department of Quantitative Methods
Universidad Pontificia Comillas – Madrid, Spain

### Abstract

The Inca Empire, which flourished in South America's Andes from the 13th to the 16th century AD, succeed agriculturally. This article discusses the Incas' terrace farming innovations and compares these constructions with other similar concepts around time and geographies. The Incas overcame the harsh Andean environment by building terraced farms, assuring a steady food supply for their large empire. It describes Inca terraces' construction, design, and ecological advantages. Inca agriculture relied on excellent irrigation infrastructure and crop diversification. There is a vast literature that emphasizes Inca terrace farming's sustainability and applicability to modern agricultural issues. The Incas' environmentally harmonious land usage, soil erosion prevention, and water resource management may help contemporary agriculture adapt to climate change. By studying the Inca Empire's agricultural accomplishments, we learn how ancient civilizations adapted to their circumstances, used natural resources effectively, and sustained agriculture for millennia. This understanding affects global food production systems as we face land degradation, climate change, and sustainable farming. Inca terrace farming is a sustainable and innovative food production method that is still relevant today. We can make global food production more sustainable and resilient by studying their past and applying their ideas to modern agriculture.

**Keywords:** Inca Empire, Terrace farming, Agricultural innovation, Food production, ancient civilizations, Crop diversity

## 1. INTRODUCTION

Because the Incas maintained control over four distinct climatic zones, this South American ancient nation was able to cultivate a wide variety of crops. Their inhabitants of the Andes were mostly vegetarian, but when they could get their hands on it, they did augment their diet with camelid meat and fish. The Incas created a massive agricultural system in which the crops and herds of conquered peoples were appropriated by the Incas, and the people of the conquered lands were forced to labour on state-owned farms at regular intervals.

The huge network of storage facilities that the Inca created as an insurance policy against times of drought and catastrophe was one of the more favourable benefits of Inca rule for the local people. In addition, food items were often given away as presents by Inca officials, with the expectation that doing so would boost the favour of the Inca monarchs.

At the most basic level, each family unit was responsible for the production of its own food. Farmland was often held by an extended family network, known as an *ayllu* (the Quechua words in *italic*, thereafter), rather than by individual households. In an ideal world, an *ayllu* would have at least some lands in both the highlands and the lower, more temperate regions of the country. This would allow for the cultivation of a wider variety of crops. For instance, the lowlands are the only suitable growing conditions for coca bushes for their leaves, but the highlands are ideal for grazing and may support the cultivation of potatoes and maize. For newlyweds to be able to provide for themselves, the newlyweds' *ayllu* would gift them a tutu, which is a tract of land suitable for growing maize and measuring around 1.5 acres. Additionally, since it was the couple's first child, they were entitled to an additional half *Tupu*. If the owner of property passed away without leaving any heirs, the land was given back to the ayllu so that it may be redistributed at a later time.

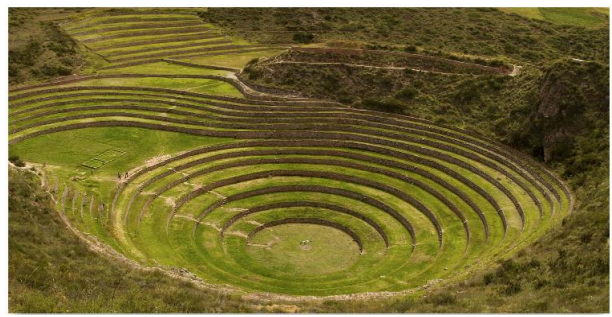

**Figure 1:** Inca Agriculture Terracing in Cusco, Peru

Simple tools such as a hoe, clod breaker, and foot plough known as the *chakitaqlla* were used to cultivate the land. The *chakitaqlla* consisted of a pointed wooden or bronze pole that was driven into the ground by putting one's foot on a horizontal bar. Historically, hoe blades were often fashioned from cobble stones that had been sharpened. Farmers worked in small teams of seven or eight, frequently singing while they worked, with the men hoeing and the women following after, breaking up clods and spreading seeds. Agriculture was a communal endeavour, and farmers worked together to cultivate their land. In the meanwhile, the family's youngsters and young adults were tasked with the responsibility of caring for the camel herd.

Maize, coca leaves, beans, grains, potatoes, sweet potatoes, *olluco*, oca, *mashua*, pepper, tomatoes, peanuts, cashews, squash, cucumber, quinoa, gourd, cotton, *tarwi*, carob, *chirimoya*, *lucuma*, guayabo, and avocado were among the crops that were farmed across the Inca Empire. The majority of the livestock consisted of herds of llamas and alpacas. Because they supplied wool, meat, leather, transportable wealth, transportation, and particularly transportation for the army, these animals were essential to many areas of Andean culture. They were also often sacrificed in religious rites. A state census was carried out every November to account for every animal in the state's many herds, which ranged in size from a few hundred to tens of thousands of individuals depending on the size of the herd.

The Incas were ambitious farmers, and in order to maximize agricultural productivity, they altered the environment by terracing it, canalizing it, and establishing irrigation networks. They also often drained marshes in order to make them ideal for cultivation. In addition, the Incas were well aware of the benefits of rotating their crops on a regular basis. Furthermore, they fertilized the soil with dried llama dung, guano, or fish heads if any of these things were accessible to them at the time. In spite of this, the climate in the Andes was notoriously unforgiving, and it was not uncommon for there to be periodic outbreaks of illness as well as floods, droughts, and other severe weather conditions. In such circumstances, the Incas' skill in the storing of food came into its own.

## 1.1. Food storage

Storehouses known as tambos (adapted to Spanish from Quechua *tampu*) or *qollqa* were constructed in the tens of thousands around the empire. These structures were often laid out in neat rows and located in close proximity to population centres, big estates, and wayside stops.

Foodstuffs and other items were kept in these *qollqa*. The *quipu*, a recording instrument comprised of threads and knots, was used by state authorities to maintain accurate stock statistics. *Qollqa* or tambos were structures made of stone with a single chamber. They might be round or rectangular in shape, and their construction was extremely consistent. The *qollqa* were built to maximize the amount of time that perishable items could be stored inside of them and were positioned on slopes so that they could take use of the natural winds. Because the inside was kept as cold and dry as possible with the use of drainage canals, gravel flooring, and ventilation in both the floor and ceiling, regular items could be stored for up to two years, while freeze-dried meals could be preserved for up to four years. Archaeologists have determined that the most often preserved foodstuffs in qollqa were maize, potatoes, and quinoa. These three items were also the most prevalent. Corn and coca from these stocks were routinely distributed to the general populace by kings who were seeking favour, as well as at times of widespread crop disasters.

## 1.2. Agriculture and Religion

The Incas considered their rituals, chants, and sacrifices to be an essential component of their agricultural practices. Llamas and guinea pigs were offered as sacrifices during these types of rites, and chicha beer was poured into the ground and into the water near rivers and springs in an effort to curry favour with the gods and the natural forces. In addition, because of the often-harsh climate of the Andes, agriculture was seen as a type of warfare. As a result, the Incas approached farming with weapons in their hands and prayers on their lips, as the historian T. N. D'Altroy so eloquently phrased it: "The Incas approached farming with weapons in their hands and prayers on their lips." Additionally, the Inca city of Cuzco was home to a great number of holy lands. The harvest from these was utilized as gifts at temples, and a specific area was set aside for the ceremonial planting of the first corn crop of the year. It was here, in the month of August, that the Inca ruler used a golden plough to perform the ritual of ceremonially tilling the first soil of the year. The holy *Coricancha*, or Temple of the Sun, one of the most important pieces of Inca architecture in the city of Cusco (Peru) which housed a shrine to the Inca sun deity Inti, had a life-size cornfield constructed entirely of gold and silver, replete with precious metal animals and insects. The cornfield was the size of an entire city block. After the Incas had successfully conquered a region, they would divide the land and cattle into three unequal parts: one for the state religion, one for the monarch, and one for the people who lived in the area. Alternately, as the tax was often taken in the form of labour (*mita*), farmers were transported to work the fields of the Inca rulers or to assist in the construction of roads and other huge structures as part of other state projects. The agricultural food that was grown on the farmers' own property was, for the most part, not affected, and the farmers were also given permission to cultivate small plots of land beside the state farms while they were doing their *mita*.

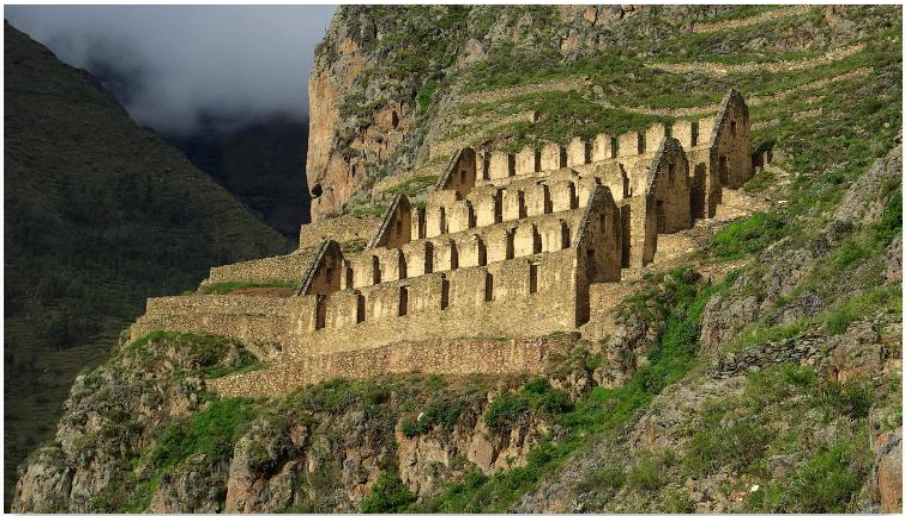

**Figure 2:** Inca *Qollqa*, Cusco – Peru

## 1.3. Inca food and drink

The Incas ate their two primary meals of the day while sat on the ground and without the use of tables. The first meal of the day occurred in the early morning, and the second occurred in the late evening. Meat, including camelid, duck, guinea pig, and wild game such as deer and the vizcacha mouse, was considered to be of such high value that it was only eaten on ceremonial occasions by the Inca. As a result, the average Inca diet consisted mostly of plant-based foods. Freeze-dried beef, also known as *charki*, was a popular dietary option for those who were often on the move. A cereal produced from quinoa was the primary source of nutrition, while seafood, generally prepared as stews, was consumed in coastal areas. Inca fisherman would use tiny boats made of reeds in the hopes of catching shellfish, anchovies, sardines, tuna, and sea bass. There were sour cherries, custard apples, elderberries, cactus fruits, pineapples, and a variety of bananas among the wild fruits that were accessible.

Stone or clay stoves were used to cook meals over fires made from wood or llama dung. The majority of the food was either boiled or roasted during this process. Corn was prepared by baking it into little cakes or toasting it, and popcorn was reserved for rare occasions. Another essential staple food was potatoes, which could be preserved in the form of *chuño* by air-drying, freeze-drying, or all these methods. Along with the grains quinoa and *cañihua*, which were very significant, the tubers oca, *mashua*, and maca were also quite important. Pounding the grains with a pestle or between two stone mortars was the traditional method of grain preparation. The addition of herbs and spices, particularly chili peppers, resulted in an increase in the intensity of the flavours. *Chicha*, a fermented beer-like drink that was created by women by chewing maize or other plants and then leaving the pulp to ferment for many days, was by far the most popular alcoholic beverage. *Chicha* had just a trace amount of alcohol.

## 2. LITERATURE REVIEW

This review will explore the key themes and findings of the selected references to gain a comprehensive understanding of how the Inca Empire achieved agricultural sustainability through various aspects of the organization of its society, always with the central objective of obtaining enough food to sustain its population and making that sustainable and planned food production capacity a mechanism for negotiation with other Andean peoples. For the focus of this article, terraced agriculture and agricultural innovation in the Inca Empire represent a remarkable example of sustainable food production practices that have captivated the attention of scholars from various disciplines.

### 2. 1. On terraced agriculture and agricultural innovation

Terraced agriculture played a central role in the agricultural system of the Inca Empire, as emphasized by **Dillehay (2002)** and **Kolata (1993, 1996)**. The construction of terraces allowed the Incas to take advantage of the challenging mountainous terrain for agricultural purposes. These terraces were not merely structures for cultivation, but also integral components of a broader agricultural strategy that promoted sustainability and crop diversification as a means of reducing risk of scarcity and famine. Recent research by **Castro et al (2019)** and **Sieczkowska et al (2022)** is clear evidence of how advanced the Incas were in terms of practical engineering and feasible waterways to maximize production, while maintaining reasonable methods to reduce the risks of adverse drought and flood situations through a rather heuristic approach.

### 2. 2. On the ritual and cultural aspects

**Bauer and Stanish (2001)** emphasize the ritual and pilgrimage aspects of Inca agriculture, especially in places like the Islands of the Sun and the Moon in Lake Titicaca. These practices may have referred to the mystical origin of the birth of their civilization on Lake Titicaca, such as the legends of *Manco Capac* and *Mama Ocllo*, popularly mentioned as the founders of the Inca empire. They suggest that these practices were closely connected with agricultural activities, reflecting the spiritual connection the Incas had with their farmland. It is also highly probable that the long droughts and terrible floods that seem to have affected the central area of the South American Pacific coast around the year 1200 AD were the trigger for the search for production methods and  social organizations that were much more effective in defeating, or at least alleviating, the climatic adversities suffered in those years and which led to the disappearance or decline of important pre-Inca cultures such as the *Huari*, *Tiahuanaco*, *Mochica* and *Nazca*, among others.  Understanding the cultural dimensions of Inca agriculture is therefore crucial to appreciating the food, population and social sustainability achieved.

### 2. 3. On gender and cultural negotiations

The work of **Browman (2003)** highlights the role of women in the foundation of Americanist archaeology, which can shed light on how gender dynamics influenced the agricultural practices of the Inca Empire. Exploring the roles of men and women in terrace agriculture and food production can offer information about the sustainability of Inca agriculture, particularly in relation to the size and altitude for the design and maintenance of the terraces, as well as for the division of labour, in terms of sowing, caring for the crops, harvesting the crops and properly storing the seeds and surplus production. This is clearly a subject open to deep and challenging research.

### 2. 4. On economic systems and markets

The ideas of **Turner (1974, 1978)** explore the economic aspects of the agricultural system of the Inca Empire. He distinguishes between supply on demand and the mechanisms of supply and demand within the Inca economy. Understanding the economic foundations of Inca agriculture can provide a valuable context for assessing its sustainability. In our opinion, the Inca economy was a clear example of a barter market under hierarchical cooperation, in which not only goods were traded, but also information and knowledge. In this sense, the Inca economic system, despite not having a currency or private rights to the means of production, particularly the land, was able to provide a sufficient base of goods, mainly food, tools, textiles and housing, within a centralized planning system, but based on cooperation between the parties involved. I maintain that the acceptance of the common people of the Inca Empire with respect to their authorities, especially the Inca nobility, was very probably since this social class shared practical knowledge regarding the best cultivation of food and the management of agricultural engineering works. That is to say, in this barter economy, the Inca nobility, made up largely of teachers (*amautas*) and planners (*quipucamayocs*), could get the people to work their land not only with a production surplus as a kind of salary paid in kind, but also for the knowledge and information that this educated class transmitted to other social sectors of the Inca Empire.

## 2. 5. On political and social structures

The ideas of **D'Altroy (2002)**, I**sbell (2008)** and **Stanish (2003)** shed light on the political and social structures that underpinned sustainable agriculture in the Inca Empire. Examining how basic finances, wealth finances and storage were managed within the Inca political economy offers insight into the governance of agriculture and food production, as we have noted above.

## 2. 6. On historical context and cultural identity

The historical account of **Sarmiento de Gamboa (2007)** provides a broader historical context for the Inca Empire, including its agricultural innovations. Understanding the historical trajectory of the empire is essential to understanding how and why these agricultural practices developed. The work of **Topic (2009)** investigates the cosmological basis of authority and identity in the Andes, offering a perspective on how cultural beliefs and cosmology influenced agricultural practices. This dimension is crucial to understanding the holistic approach to sustainability in Inca agriculture. This approach, in considering the divinity of the Inca as the son of the Sun god (*Inti*), is not unlike that of many other cultures, especially in Asia (China, Japan) and Africa (Egypt).

## 3.  RESEARCH METHODOLOGY

Investigating and analysing how the hierarchical cooperation and barter economy of the Inca Empire functioned is one of the primaries focuses of this historical and anthropological study. This study's objective is to gain an understanding of the agricultural practices that the Incas employed, as well as the commerce that occurred both within their empire and with the cultures that were located in close proximity to it, and the role that hierarchical collaboration played in maintaining the stability of the empire's economic and social structure.

### 3.1. Research Questions

1. In order to sustain their economy based on barter, what were the primary agricultural practices that the Incas used?

The Incas were extremely adept in agricultural operations, which played an important part in maintaining their barter economy. The Incas governed one of the most extensive empires in pre-Columbian America, and their empire was one of the largest in the region. They were able to maintain a huge population because to the inventive and complex farming practices that they used, which also made it easier for them to trade and communicate with people from other parts of the empire.

Farming on terraces, the Incas were the undisputed masters of terrace farming, a technique that included sculpting level platforms into steep hillsides to produce cultivable ground for agriculture. The construction of these terraces served several functions, including the prevention of soil erosion, the maximization of water retention, and the provision of appropriate conditions for the cultivation of crops at varying elevations. Irrigation systems: The Incas created complex irrigation systems to channel water from natural sources like as rivers and springs to agricultural fields in areas that received a low amount of precipitation. These systems were used in regions that were home to Inca cities. Crops were able to flourish in desert locations because to the innovative construction of canals and aqueducts, which ensured a constant supply of water for the crops and allowed them to survive.

The Incas used a kind of agriculture known as *Waru Waru*, which included the use of elevated beds. In order to do this, high platforms with ditches on each side had to be constructed. The raised beds served to shield the crops from frost and pests, while the ditches assisted in regulating the flow of water and gave the crops with extra moisture.

Crop rotation was an important technique for the Incas, since it helped them preserve soil fertility and stop the loss of nutrients over time. They might maximize the efficiency with which nutrients are used and decrease the likelihood that the soil would become depleted if they rotated the sorts of crops planted on a given piece of land throughout the year. Potatoes that have been freeze-dried the Incas came up with a one-of-a-kind way of preserving potatoes, which were one of their major crops. They did this by exposing the potatoes to the subfreezing conditions that prevailed at the high Andean heights. This caused the potatoes to freeze-dry. They were able to store enormous amounts of potatoes for lengthy periods of time, which ensured that they would have access to food even during times of famine.

Thanks to multiple agricultural zones, since the geography of Inca Empire comprised a varied range of temperatures and ecosystems, ranging from high-altitude alpine areas to coastal deserts. This allowed the Incas to cultivate crops in a variety of environments. They did this by establishing a number of agricultural zones, collectively referred to as a "vertical archipelago," in which different kinds of plants were grown at varying elevations according to the specific temperature and humidity needs of each kind.

Another key factor was the agricultural work and its organization. The Incas used a method known as "*mita*" to organize their communal work in the agricultural sector. Communities were organized to participate in agricultural projects that were funded by the state. These projects included the maintenance of terraces and irrigation systems. Due to the combined efforts of everyone involved, more productive farming techniques were able to be implemented on a wider scale.

2. How did the Inca barter economy make it easier for commerce and the exchange of resources across the many areas of the empire and with the cultures that were nearby?

The Inca barter economy was crucial in promoting commerce and the exchange of resources, not only across the provinces of the empire but also with the cultures that were in close proximity to it. The Incas built one of the most extensive empires in pre-Columbian America, and as a result, they were able to develop a commerce network that was both well-organized and efficient. This network enabled them to trade a broad variety of products and resources. Their extensive road system, which linked different sections of the empire and was known as the "Inca Road" (*Qapac Ñam*) made it possible for the free flow of commodities, people, and knowledge across the empire.

The economy of the empire was based on barter, and it functioned according to a system that included redistribution and reciprocity. Local communities were responsible for the production of excess agricultural items, textiles, metals, and handicrafts, which were later gathered in administrative centres and kept there. After that, the excess items were reallocated to regions that were suffering from a lack, so guaranteeing that an equal distribution was achieved and providing assistance to places with less favourable agricultural circumstances. This component of the Inca barter economy's redistributive practices helped to create social cohesiveness and contributed to the empire's overall stability.

The Incas engaged in significant commerce not just within their own society but also with the cultures that bordered them as well as with countries beyond their limits. Because of their geographical position in the Andean area, they had access to a broad array of resources from diverse ecological zones. Some of these resources were coca leaves, cotton, and tropical fruits. The Incas had a sophisticated system of bartering that allowed them to acquire many valuable resources in return for agricultural goods, textiles, and precious metals. The Inca barter economy did not rely on the use of money but rather on a system of standardized units of trade known as *khipu* (often written as *quipu*) which were complicated threads with knots of various colours and places expressing numeric values. These *khipu* were used as a complex accounting tool by the Incas, which allowed them to keep track of commercial transactions as well as the allocation of resources.

The function of the state's central authority was essential to the process of regulating commerce and guaranteeing equitable transactions. Along the Inca Road, merchants and traders travelled under the protection and supervision of state authorities, which decreased the likelihood of being robbed or attacked by bandits and improved the safety of trade routes.

## 4.  COMPARISON ANALYSIS AND DISCUSSION

Terrace farming, a legendary agricultural practice, has been instrumental in enabling cultivation on steep terrains across various continents. This technique not only maximizes arable land but also plays a crucial role in soil conservation, water management, and adaptation to diverse climatic conditions. Below is a comprehensive overview of terrace farming implementations in South America, Asia, Africa, and Europe, supported by scholarly references.

### 4.1 In South America: The Andean *Andenes*

In the Andean regions of South America, particularly within the Inca Empire, terrace farming—locally known as *andenes*—was extensively developed to adapt agriculture to mountainous landscapes. These terraces transformed steep slopes into productive agricultural lands, facilitating the cultivation of crops like potatoes and maize. The construction involved intricate stone walls and sophisticated drainage systems, reflecting advanced engineering skills. *Andenes* not only increased arable land but also mitigated soil erosion and optimized water usage, showcasing a sustainable approach to high-altitude farming. Many of these terraces remain functional today, underscoring their enduring utility.

### 4.2 Asia: Rice Terraces and Sustainable Practices

Asia boasts some of the most iconic examples of terrace farming, particularly in rice cultivation. In countries like China, the Philippines, and Nepal, rice terraces have been integral to food production and cultural heritage. The Banaue Rice Terraces in the Philippines, often referred to as the "Eighth Wonder of the World", exemplify ancient ingenuity in creating sustainable agricultural systems that harmonize with the environment. These terraces are designed to follow the natural contours of the mountains, reducing soil erosion and managing water resources efficiently. However, modern challenges such as labour shortages, urbanization, and climate change have led to the abandonment of some terrace systems, posing risks to food security and cultural landscapes.

### 4.3 Africa: Terracing in Arid and Semi-Arid Regions

In Africa, terrace farming has been a vital adaptation strategy in arid and semi-arid regions. In the Ethiopian Highlands, farmers have long employed terracing to combat soil erosion and enhance water retention, thereby improving agricultural productivity. Similarly, in the Southern Levant, encompassing parts of modern-day Israel, Jordan, and Egypt, ancient communities constructed extensive terrace systems to harness scarce water resources and cultivate crops in desert environments. These practices not only supported food production but also contributed to the stabilization of fragile ecosystems.

## 4.4 Europe: Mediterranean Terraces and Cultural Landscapes

Europe's Mediterranean region has a rich history of terrace farming, particularly in countries like Italy, Spain, and Greece. These terraces have been instrumental in cultivating olives, grapes, and other crops on hilly terrains. The dry-stone walls characteristic of Mediterranean terraces not only support agriculture but also represent significant cultural heritage. However, socio-economic changes and urban migration have led to the neglect and abandonment of many terrace systems, resulting in increased soil erosion and loss of biodiversity. Efforts are underway to restore and preserve these terraces, recognizing their ecological and cultural importance.

Tables 1 to 3, as seen below, summarise the origin, time and the characteristics of farming terraces during the last eighteen centuries. Inca terrace farming was characterized by high-altitude stone constructions with deep drainage systems, enabling cultivation of diverse crops such as potato, maize, quinoa, oca, and olluco across up to 20 ecological zones. In contrast, Asian terraces—mainly from China and Southeast Asia—were lower in altitude, primarily rice-based, and relied on gravity-fed water systems. Despite regional differences, both systems aimed to control soil erosion, manage water efficiently, and stabilize slopes. Socially, Inca maintenance was state-directed (e.g., mit'a labor), while Asian systems were more community-based. Mediterranean, Ethiopian, and other global terraces also adapted to their specific environmental conditions, demonstrating how ancient societies used terraces not only for agriculture but for resilience and sustainability across diverse geographies.

Peer Reviewed
Multidisciplinary
International

Table 1. Global History of Terrace Farming

| Region / Culture | Approximate Period | Key Characteristics |
|---|---|---|
| Inca Civilization (Andes) | 13th–16th centuries CE | Stone terraces with drainage channels, diverse microclimates, transported topsoil. |
| China (Han Dynasty onward) | Since 2nd century BCE | Rice terraces in mountainous zones (e.g., Longji); water use by gravity. |
| Southeast Asia | 1st–10th centuries CE | Wet rice terraces with communal irrigation systems. |
| Ethiopia and Horn of Africa | Since 1st millennium CE | Soil conservation terraces for cereals in semi-arid zones. |
| Mediterranean (Greece, Rome) | Classical Antiquity | Dry stone terraces for vineyards and olive trees on hilly Mediterranean slopes. |

Table 2. Similarities Between Inca and Asian Terrace Farming

| Common Feature | Inca Empire | Asia (Mainly China and Southeast Asia) |
|---|---|---|
| Soil erosion control | Prevented landslides, retained soil on slopes | Same goal: reduce runoff and preserve arable land |
| Hydraulic engineering | Complex drainage and vertical water channels | Gravity-fed irrigation systems with canals |
| Social organization | State-managed maintenance (e.g., mit'a labor) | Village or community-based cooperative maintenance |

Table 3. Key Differences Between Inca and Asian Terrace Farming

| Characteristic | Incas | Asia |
|---|---|---|
| Altitude | Very high (2,500–4,000 m a.s.l.) | Moderate (typically 500–1,500 m a.s.l.) |
| Main crops | Potato, maize, quinoa, oca, olluco | Primarily rice |
| Construction materials | Finely carved stone, imported soil | Soil-based terraces, sometimes with dry stone or wood |
| Water technology | Multilevel irrigation and deep drainage systems | Flood irrigation, gravity-fed basins |
| Ecological diversity | Managed up to 20 ecological floors | Generally monoculture or dual-crop zones |

As per our discussion, we may say that terrace farming and agricultural innovation in the Inca Empire were multifaceted and deeply integrated into the culture, economy, and society of the time. The sustainable approach to food production in the Inca Empire was a result of complex interactions between agricultural techniques, cultural beliefs, and economic structures. Future research should continue to explore these themes and their relevance for contemporary sustainability challenges in agriculture.

For the Incas, the Altiplano saw a mind-boggling expansion of water and diverse types of water across the board with other rural methods. In addition, it was often used in different regions of South America that lacked usual leaking or had short however extreme rainy seasons followed by extended dry ones. The process of strip drainage involves building narrow channels all the way across the field and storing dirt on the edges and in the middle between the channels to create edges on which to plant crops.

Even though they were originally studied by **Flores and Paz (1983)** and then once again by **Rosas (1986)**, as opposed to the fields that had been furrowed, their use has never been abandoned, and they are still widely used today. Their widths vary from around 20 to 60 meters, and they are made up of numerically normal curved dejections that have been hacked into the plain or reconstructed using standard concavities as their source material. The edges of the *qocha* are etched with a spiral pattern of wrinkles on a few levels, and the harvests are planted on the delicate in the spaces in between. The *qocha* are linked to one another by a network of canals that is laid out in a pattern that resembles a string of dots. The precipitation that falls from the upper bowl fan that is located behind them is where the water first starts to form. The water levels in the *qocha* are managed by an organization that consists of embankments and repositories. These are opened and closed depending on the amount of precipitation that has fallen as well as other factors, and the flood is then directed into streams and streams that flow into Lake Titicaca. These artifacts, much like the furrowed fields, are best observed in order to reduce the potential for danger. Numerous *qocha* may be used as reservoirs to retain water for use during periods of drought and to irrigate plants during periods of abundant precipitation. The bottom may be used to store water, and the sides can be put to productive use by growing plants along them. According to Flores, they were around at the same time as the possible founding of the *Sukka Qolla*.

## 5. CONCLUSION

Terrace farming stands as evidence to human ingenuity in adapting agriculture to challenging environments. Across continents, these systems have enabled sustainable food production, conservation of natural resources, and preservation of cultural landscapes. Contemporary challenges necessitate renewed focus on maintaining and revitalizing terrace farming practices to ensure their benefits for future generations.The study of terrace farming as a remarkable example of agricultural innovation in the Inca Empire sheds light on a unique and sustainable method to food production that continues to attract scholars as well as supporters. The Incas were able to effectively harness the difficult mountainous terrain of the Andes by using inventive tactics such as terraced landscapes and advanced farming techniques. This not only ensured the Incas' ability to provide for their own food needs, but it also paved the way for the expansion of their empire. This approach to sustainability was intricately laced into the cultural, economic, and social fabric of Inca civilization, as is evident from the ceremonies, community customs, and economic systems that the Inca used.

The need of adaptation and resourcefulness in the face of geographical limits is shown by the Inca Empire's mastery of terrace farming and agricultural innovation. Their complexly terraced fields, along with a network of canals and irrigation, enabled them to maintain soil fertility and water management in an effective manner, both of which are essential components of environmentally responsible agriculture. Additionally, the economic and social systems design by the Incas, relying on agriculture, barter and hierarchical cooperation, was also coupled with the Incas' veneration for the land and their cosmological beliefs highlighted the holiness of their agricultural techniques, which helped to reinforce a harmonious interaction between people and their natural surroundings.

The lessons that may be learned from the agricultural techniques of the Inca Empire continue to be important even as we face modern issues in agriculture and environmental sustainability. The Incas established resilient agricultural systems via the use of terrace farming and other inventive practices that provide insight into how such systems might be developed and maintained through time. This historical example serves as a monument to the everlasting importance of environmentally responsible farming techniques, demonstrating the possibility for contemporary cultures to take inspiration from the past as they seek answers to the food production and environmental concerns of today.

**Author's Declaration**

**Prof. Luis-Felipe Arizmendi**